\documentclass{svjour3}
\usepackage[figuresright]{rotating}

\begin{document}

\title{Chaotic Spiral Galaxies}

\author{G. Contopoulos  \and  M. Harsoula}
\institute{G. Contopoulos \at
              Research Center for Astronomy,
           Academy of Athens, Soranou Efesiou 4, GR-115 27 Athens, Greece \\
              \email{gcontop@academyofathens.gr}
           \and
          M. Harsoula \at
              Research Center for Astronomy,Academy of Athens, Soranou Efesiou 4, GR-115 27 Athens, Greece\\
              \email{mharsoul@academyofathens.gr} }

\date{Received: date / Accepted: date}
\maketitle

\begin{abstract}
 We study the role of asymptotic curves in supporting the spiral
 structure of a N-body model simulating a barred spiral galaxy. Chaotic orbits with
 initial conditions on the unstable asymptotic curves of the main
 unstable periodic orbits follow the shape of
 the periodic orbits for an initial interval of time and then they are
 diffused outwards supporting the spiral structure of the galaxy. Chaotic orbits
 having small deviations from the unstable periodic orbits, stay close and along
 the corresponding unstable asymptotic manifolds, supporting the spiral structure for
more than 10 rotations of the bar.
 Chaotic orbits of different Jacobi constants support
 different parts of the spiral structure. We also study the diffusion rate of chaotic orbits outwards and
 find that chaotic orbits that support the outer parts of the galaxy
 are diffused outwards more slowly than the orbits supporting the
 inner parts of the spiral structure.

\end{abstract}

\maketitle

\section{Introduction}

It is well known that the spiral arms of galaxies are density waves.
This means that the spiral arms are not always composed of the same
matter but they only represent the maxima of the density along every
circle around the center. The stars passing through the spiral arms
stay longer close to them, thus producing an increase of the local
density.

The linear theory  of spiral density waves assumes that the
potential V, the density $\rho$ (or the surface density $\sigma$)
and the distribution function \textit{f} (phase space density) have
small deviations from their axisymmetric values
($V_0$,$\rho_0$,$f_0$).

This linear theory was initiated by Lindblad $(1940,1942)$ but it
was developed in its modern form by Lin and Shu $(1964, 1966)$,
Kalnajs $(1971)$,  Lynden-Bell and Kalnajs $(1972)$ Toomre
$(1977)$,and others.

However the deviations near the main resonances of a galaxy (inner
and outer Lindblad resonances and corotation) are large, thus near
these resonances a nonlinear theory is necessary (Contopoulos
$1970,1973,1975$).

In both the linear and the nonlinear theory, whenever the
perturbation is relatively small, chaos is unimportant. Although
some chaotic orbits appear near all the unstable periodic orbits,
their proportion and their effects are small. This is the case of
most normal spirals, where the density perturbations are of order
$2-10\%$ of the axisymmetric background.

On the other hand in barred galaxies the perturbations are large, of
the order of $50-100\%$. In such cases the chaotic orbits play an
important role in the dynamics of the galaxies. The main chaotic
effects in a galaxy appear near corotation.

It is well known now that chaos is generated by the overlapping of
resonances (Contopoulos $1966$, Rosenbluth et al. $1966$, Chirikov
$1979$). In the region near corotation there are many resonances
between the angular velocities of the stars $(\Omega-\Omega_s)$ (in
the frame rotating with the angular velocity of the pattern
$\Omega_s$) and the epicyclic frequence $\kappa$.

The ratio of these frequencies is
\begin{equation}
q =\frac {\Omega-\Omega_s}{\kappa}
\end {equation}

These resonances are congested in a relatively small interval of
Jacobi constants $E_j$ around the Jacobi constant $E_{j0}$ at
corotation. Some important resonances in an extended region around
corotation are $ 1/q=4/1, 3/1, 2/1$ (inside corotation), $q=0$ (at
corotation) and $-2/1$,$-1/1$ (outside corotation). Chaos produced
by the interactions of resonances, extends around the envelope of
the bar and along the spiral arms beyond the end of the bar. The
first example of a chaotic orbit that fills an envelope of the bar
and the inner parts of the spiral structure was provided by Kaufmann
and Contopoulos (1996) (Fig.1).

\begin{figure}
\centering
\includegraphics[width=8.cm] {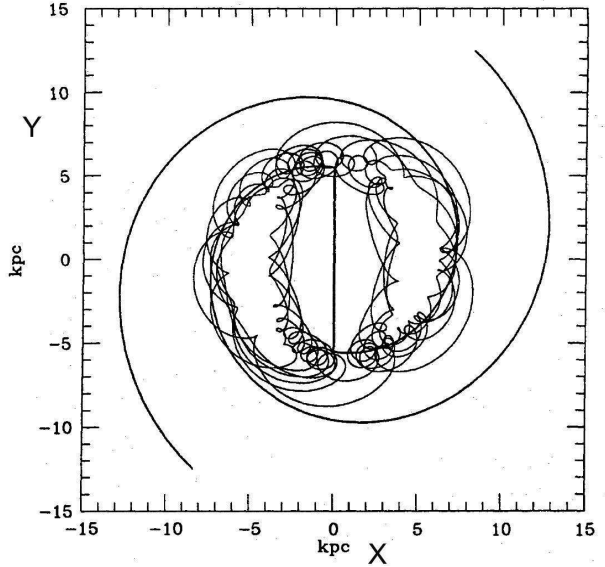}
\centering \caption{First example of a chaotic orbit in a galactic
model of NGC 3992 in \cite{Kauf1996}.} \label{fig2}
\end{figure}

However the most surprising result came from N-body simulations,
which indicated that most orbits along the spiral arms beyond the
ends of the  bar are chaotic (Voglis et al. $2006 a$).

Thus started a systematic study of the chaotic spiral arms outside
corotation in strong barred galaxies (Voglis and Stavropoulos 2005,
Voglis et al. $2006 a,b$, Romero-Gomez et al. $2006$,$2007$,
Tsoutsis et al. $2008$, Athanassoula et al.$2009 a,b$, \textbf{2010},  Harsoula et
al.$2009$, $2011$). In the present paper we present the most recent
results of this study.

\section{Chaotic Density Waves}

Although the spiral arms beyond the ends of the bars are composed of
chaotic orbits, nevertheless these spiral arms are density waves,
i.e. the density maxima are populated by different stars at every
time.

The density of the stars is maximum in areas where their velocities
are minimum. This happens mainly near the apocentres and the
pericentres of the orbits. And these apocentres and pericentres
appear close to the asymptotic manifolds of the main unstable
periodic orbits around corotation.

The most important families of unstable periodic orbits in the
corotation region are the families $PL_1$,$PL_2$ around the unstable
Lagrangian points $L_1$ and $L_2$.

In a simple model of a barred galaxy the Hamiltonian near
corotation $(r_*)$ (Contopoulos $1978$) is given by the relation
\begin{equation}
H=h_\ast+\kappa_\ast I_1+a_\ast I_1^2+2b_\ast I_1 I_2 +c_\ast I_2^2
+ A_\ast \cos 2\theta_2=h
\end{equation}
where $h$ is the Jacobi constant, $I_1$ the epicyclic action,
$I_2=J-J_\ast$ is the azimuthal action, (where $J$, $J_\ast$ are the
angular momenta of the star and of corotation) $\theta_2$ is the
azimuth of the star and $h_\ast$, $\kappa_\ast$, $a_\ast$,
 $b_\ast<0$, $c_\ast<0$, $A_\ast$ are constants. The action $I_1$ is
also  constant, because the conjugate angle \textbf{$\theta_1$}, does not
appear in the Hamiltonian. The orbits $PL_1$, $PL_2$ are represented
by the unstable equilibrium points of the system, namely when
\begin{equation}
\partial H/\partial I_2=\partial H/\partial\theta_2=0
\end{equation}
Therefore
\begin{equation}
 sin2\theta_2=0
\end{equation}
and
\begin{equation}
b_\ast I_1+c_\ast I_2=0
\end{equation}
The orbits close to the Lagrangian points $L_1$, $L_2$ have
$\theta_2=\pi/2$ and $\theta_2=3\pi /2$, (while the orbits close to
the Lagrangian points $L_4$,$L_5$ have $\theta_2=0$, and
$\theta_2=\pi$),

In the lowest approximation the orbits $PL_1$, $PL_2$ start with
$\theta_2=\pi/2$, or $\theta_2=3\pi/2,$ and
\begin{equation}
h_\ast+\kappa_\ast I_1- A_\ast =h
\end{equation}
hence
\begin{equation}
I_1=\frac{h-h_\ast +A_\ast}{\kappa_\ast}>0
\end{equation}
Therefore these orbits exist whenever $h>h_\ast -A_\ast$.

 After finding $I_1$ we can find $I_2$ from Eq.(5).

 We have (Contopoulos $ 1975$)
\begin{equation}
b_\ast =\frac{\Omega_\ast \kappa'_\ast}{r_\ast
\kappa^2_\ast},c_\ast=\frac{\Omega_\ast \Omega'_\ast}{r_\ast
\kappa^2_\ast}
\end{equation}
where $\Omega$ is the angular velocity, and $\kappa$ is the
epicyclic frequency. The subscript $\ast$ denotes the values at
corotation and the accents mean derivatives with respect to the
radius r. Thus Eq.(5) can be written
\begin{equation}
\kappa '_{\ast} I_{1} +\Omega^{'}_{\ast} I_{2} = 0
\end{equation}
We consider now a simple model of the form
\begin{equation}
V'_o = \frac{c^2}{r^\rho}
\end{equation}
where $V_o$ is the potential and $V'_o$ is the force as a function
of $r$. Then we have
\begin{equation}
\Omega = \sqrt\frac{V'_o}{r}= \frac{c}{\sqrt r^{\rho+1}}
\end{equation}
and
\begin{equation}
\kappa = \sqrt { V{''}_o + \frac{3V'_o}{r}} = \frac{c}{\sqrt
r^{\rho+1}} (3-\rho)
\end{equation}
Thus Eq.(9) gives
\begin{equation}
(3-\rho)I_1 + I_2=0
\end{equation}
and if $\rho<3$ we find that $I_2$ is negative. In particular in a
Keplerian model $\rho =2$ and the relation (13) becomes
\begin{equation}
I_1 + I_2 = 0
\end{equation}

Therefore $I_2<0$. The initial point of the orbit $PL_1$ (with
$\theta=\pi/2$ and increasing $\theta$) is $\bar{PL_1}$ inside the
corotation distance with
\begin{equation}
\Delta r_o = \frac{2\Omega_\ast}{r_\ast \kappa^2_\ast} I_2 < 0
\end{equation}
The orbit $PL_1$ is described counterclockwise (while the galaxy is
rotating clockwise.) Thus the orbit $PL_1$ intersects the
$\theta=\pi/2$ axis once more with $\Delta r_o>0$ and decreasing
$\theta$ (Fig. 2).
\begin{figure}
\centering
\includegraphics[width=6.cm] {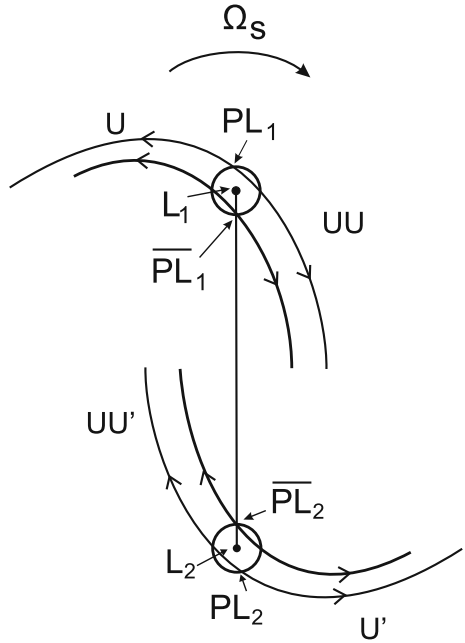}
\centering \caption{The asymptotic manifolds from the orbits $PL_1$,
$PL_2$ on the configuration plane.} \label{fig2}
\end{figure}
The asymptotic manifolds from the orbits $PL_1$, $PL_2$ on the
configuration plane are shown in Fig.2. There are two unstable
manifolds from $PL_1$ namely $U$ along a trailing spiral and $UU$
along the leading edge of the bar, and two stable manifolds, $SS$
(along a leading direction) and S (along the trailing edge of the
bar). These manifolds contain successive apocentres ($\dot{r}$=0))
of the asymptotic orbits i.e. orbits starting on this manifold.
Similar manifolds emanate from $\bar{PL_1}$, $\bar{PL_2}$,
corresponding to the pericentres of the asymptotic orbits.

The orbits starting close to $PL_1$ but not exactly on the
asymptotic manifold approach this point along the stable directions
$S$ and $SS$ and then deviate along the unstable directions $U$ and
$UU$. After a longer time these orbits form trailing spiral arms and
the envelope of the bar. The forms of the asymptotic curves
emanating from $PL_1$ and $PL_2$ are shown in Fig. 3. We see that
the curve U from $PL_1$ reaches the neighborhood of $PL_2$, making
oscillations around the asymptotic curve $SS'$ of larger and larger
amplitude as it approaches $PL_2$, coming closer and closer to the
asymptotic curves $U'$ and $UU'$ from $PL_2$, which are symmetric to
$U$ and $UU$ with respect to the center of the galaxy. Thus the
matter that starts close to $PL_1$ and moves along the asymptotic
curve $U$, after reaching the neighborhood of $PL_2$ moves very
close to the asymptotic curves $U'$ and $UU'$ and approaches again
the neighborhood of $PL_1$. In a similar way matter moves along the
asymptotic curves $UU$, $U'$ and $UU'$. Thus we have circulation of
the material along the spiral arms that lasts for a substantial
fraction of the Hubble time, until the spiral arms fade away.
\begin{figure}
\centering
\includegraphics[width=5.cm] {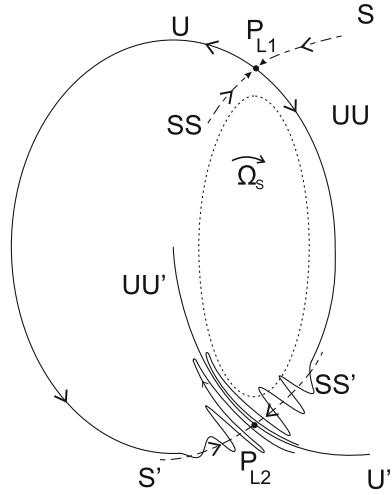}
\centering \caption{The forms of the asymptotic curves emanating
from $PL_1$ and $PL_2$.} \label{fig2}
\end{figure}
The individual orbits along the asymptotic curve $U$ or close to it,
are of the form of Fig.4. The orbits start by making some rotations
close to the periodic orbit $PL_1$ and they deviate away from
$PL_1$, reaching the neighborhood of $PL_2$. Then they proceed
either close to the spiral $U'$ (Fig.4a), or close to the envelope
of the bar $UU'$ (Fig. 4b).

The unstable asymptotic curves of the various unstable periodic
orbits for the same Jacobi constant cannot cross themselves or each
other. Thus they are obliged to follow nearly parallel paths. The
main unstable orbits inside corotation are the families $4/1$, $3/1$
and $2/1$ (inner Lindblad), while close to corotation the most
important families are $Pl_1$ and $PL_2$ whose asymptotic manifolds
are shown in Fig. 5a. On the other hand, all the manifolds of the
unstable periodic orbits corresponding to the same Jacobi constant,
are approximately parallel and contribute to the formation of spiral
arms outside corotation (Fig. 5b). This is the phenomenon of
coalescence that was described by Tsoutsis et al.(2008).
\begin{figure}
\centering
\includegraphics[width=12.cm] {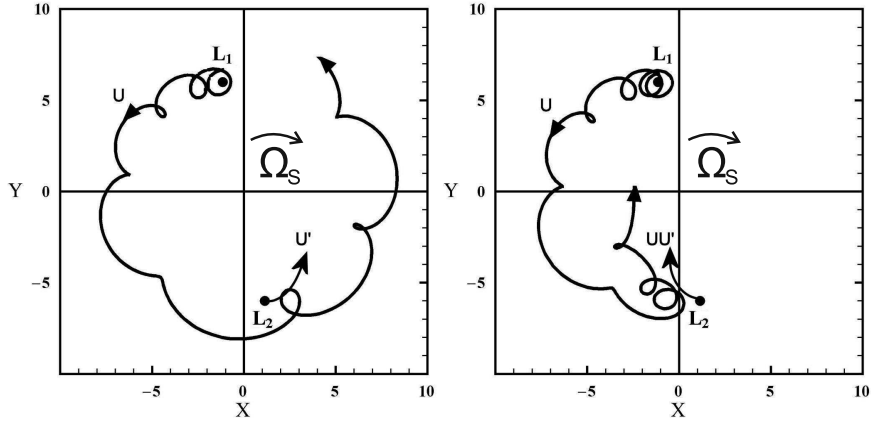}
\centering \caption{Orbits starting close to $PL_1$ (around the
Lagrangian point $L_1$) approach the Lagrangian point $L_2$ and
then deviate (a) along $U'$, or (b) along $UU'$.} \label{fig2}
\end{figure}
\begin{figure}
\centering
\includegraphics[width=6.2cm] {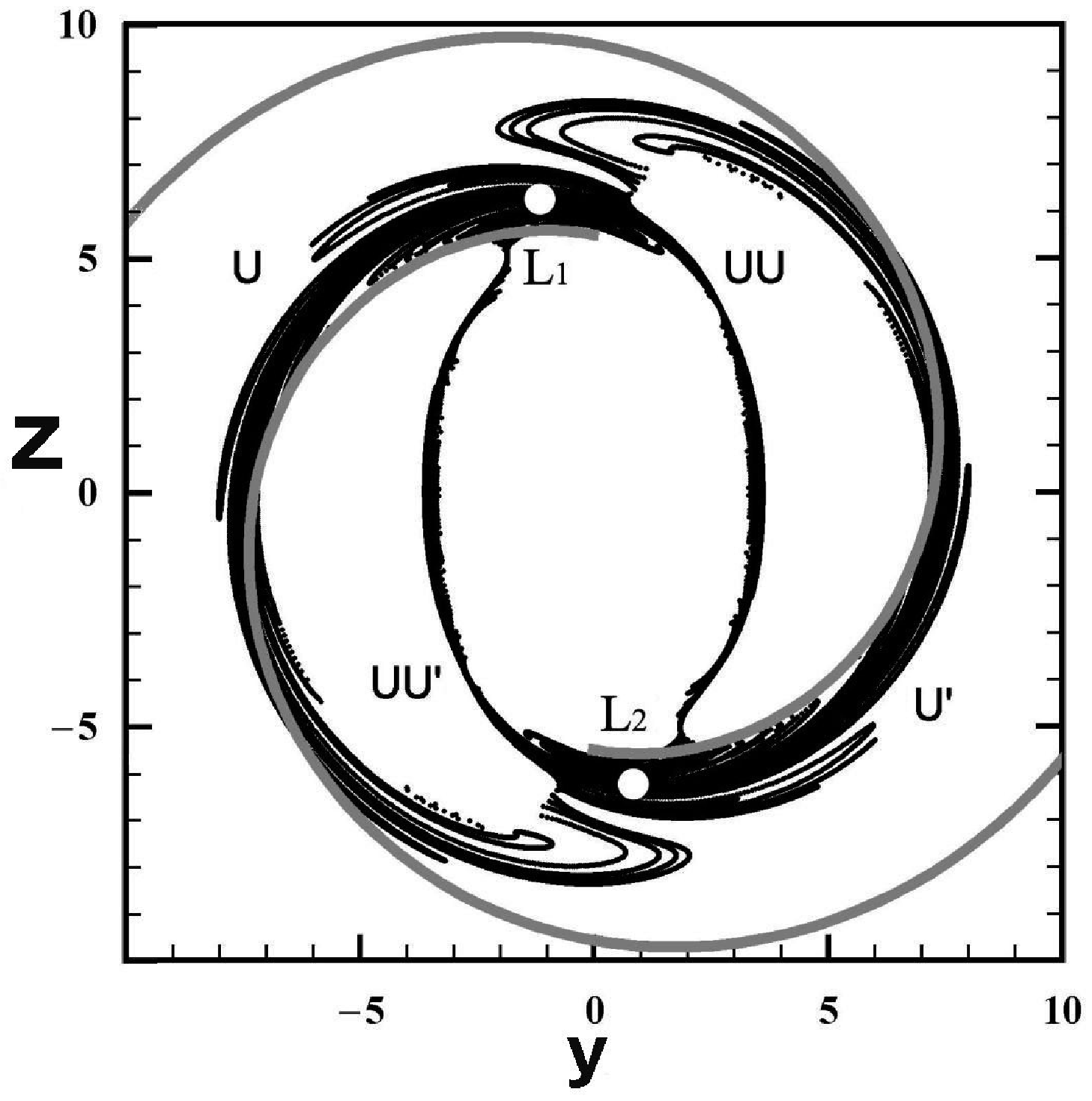}
\includegraphics[width=5.5cm] {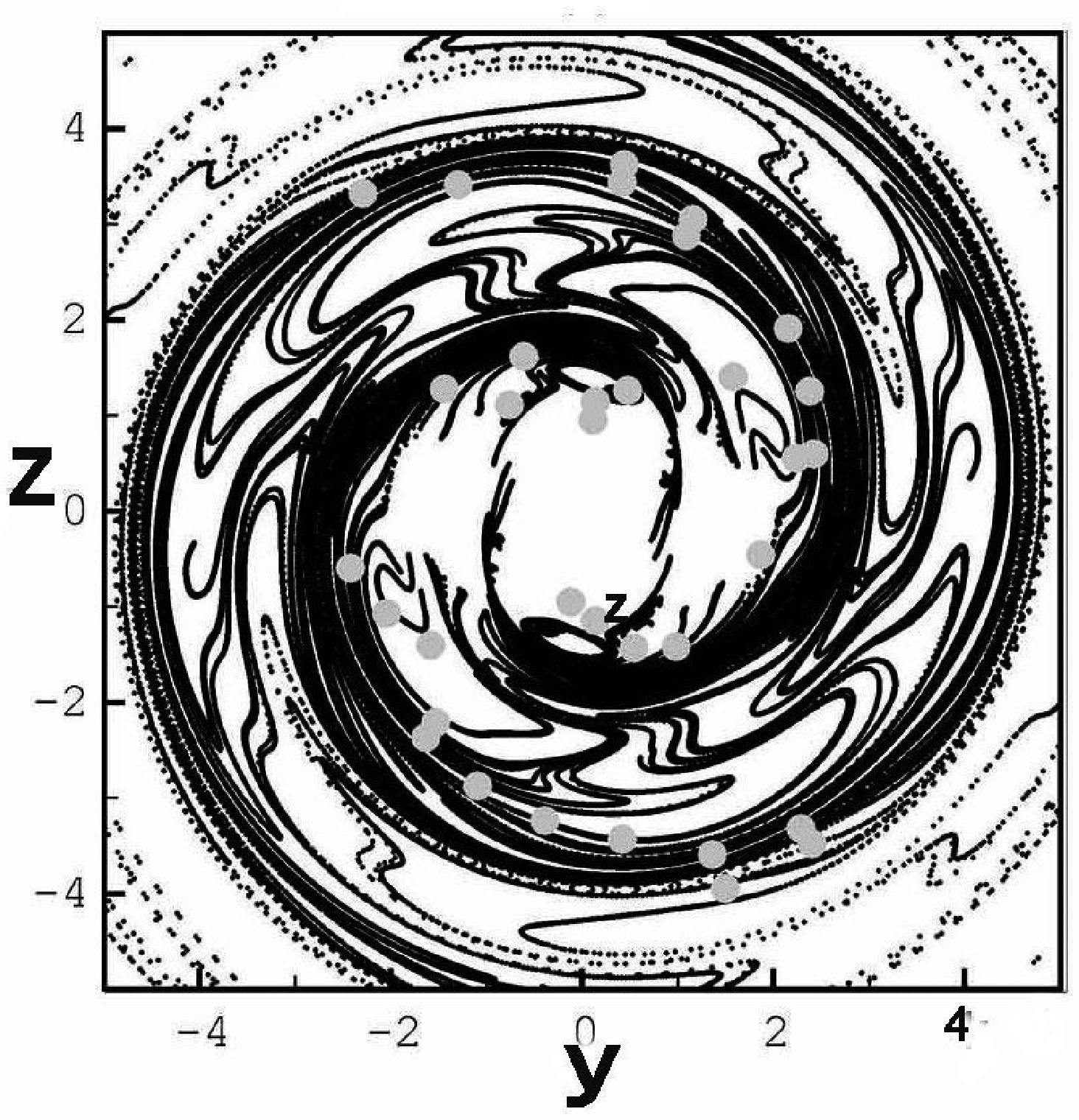}
\centering \caption{(a) The projected unstable asymptotic manifolds
from $PL_1$ and $PL_2$ in the configuration space (b) The
``coalescence'' of the invariant asymptotic manifolds from the
$PL_1$, $PL_2$, $-1//1$, $-2/1$ and $-41$ families. } \label{fig2}
\end{figure}
If we consider also similar figures for other values of the Jacobi
constant (in which case overlapping of various curves is permitted)
we see the formation of thick spiral arms. In fact the observed
spiral arms in N-body simulations of barred galaxies are very close
to the general form of the spiral arms produced by the superposition
of the various asymptotic curves.

However orbits close to the particular unstable periodic orbits
support particular parts of the spiral arms. Thus it is of interest
to study the orbits close to every resonance.

\section{Resonant orbits and diffusion times}

The planar orbits in a time independent rotating model have a fixed
Jacobi constant (that we call "energy in the rotating frame")
\begin{equation}
E_j=\frac{1}{2}u^2+V(r,\theta)-\Omega_s J
\end{equation}
where u is the velocity in the rotating frame, $V(r,\theta)$ is the
potential in polar coordinates, $J$ is the angular momentum and
$\Omega_s$ is the angular velocity of the system.

If $E_j$ is larger that the energy $E_{jo}$ at the Lagrangian points
$L_1,L_2$, the orbits inside and outside corotation can communicate.
If, however, $E_j<E_{jo}$ the orbits inside corotation cannot get
outside and those outside corotation cannot get inside.

We consider a particular N-body system that was studied by Voglis
and Stavropoulos (2005) and Voglis et al. (2006a) simulating a
barred spiral galaxy and we separate the orbits in particular
intervals $E_j\pm 10000$.

We consider first the N-body orbits that have energies in the
interval $E_j = - 1090000 \pm 10000$. In this energy level there are
almost no regular orbits at all (see Fig. 3 of Harsoula et al.
2011). For these energies the orbits can move both inside and
outside corotation. We integrate the orbits with initial conditions
outside corotation for $100 T_{hmct}$ (half-mass crossing times) and
find the distribution of their $q$ values (Fig. 6a).

The time interval $\Delta t =100 T_{hmct}$ is about one third of the
Hubble time. During that time most resonant orbits are concentrated
near the resonances $-2/1$ (outer Lindblad), $-1/1$ and $-2/3$ .
Only a few orbits have moved inside corotation $(q>0)$.

However after about one and a half Hubble time (Fig. 6b) the above
resonances have fewer stars. Some stars have escaped, but a
substantial proportion of stars have reached the inner resonances
$3/1$ and $2/1$ (inner Lindblad). These stars are trapped near these
resonances for very long times before escaping again outwards. On
the other hand stars starting inside corotation remain close to the
resonances $3/1$ and $2/1$ for more than $5$ Hubble times before
escaping outside the galaxy (without being trapped by the outer
resonances). It must be pointed out here that the 2-body relaxation
time for galaxies is of the order of $10^6-10^7$ Hubble times, a
time exceedingly longer that the diffusion time of the chaotic
orbits in our N-body model.

Then we consider the stars outside corotation for an energy interval
$E_j=-1250 000 \pm 10000$, where again only chaotic orbits exist
outside corotation. In this case the orbits with initial conditions
outside corotation cannot enter inside corotation.
\begin{figure}
\centering
\includegraphics[width=5.5cm,angle=-90] {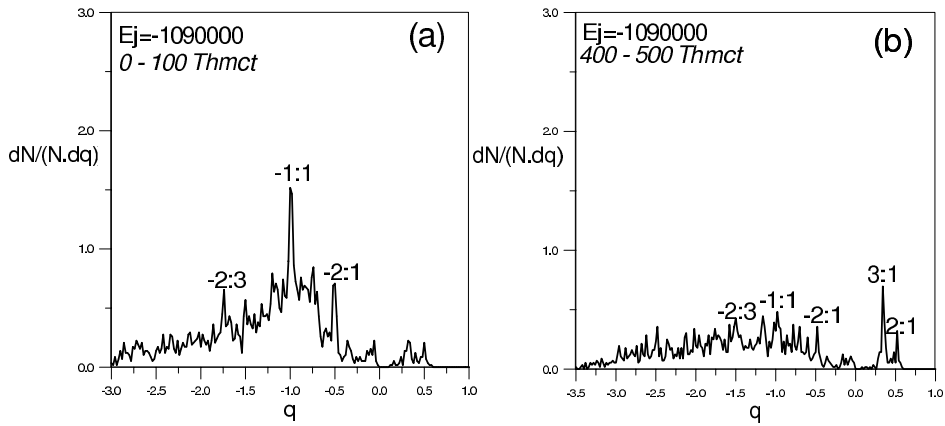}
\centering \caption{The distribution of the frequency ratios $q$ of
N-body chaotic orbits having initial conditions outside corotation
and belonging to the energy level $E_j=-1090000\pm 10000$, when the
orbits are integrated (a) for 100$T_{hmct}$ (corresponding to
$\approx$ 1/3 Hubble time) and (b) from 400$T_{hmct}$ to
500$T_{hmct}$ (corresponding to $\approx$ 1,5 Hubble time).}
\label{fig2}
\end{figure}
\begin{figure}
\centering
\includegraphics[width=5.5cm,angle=-90] {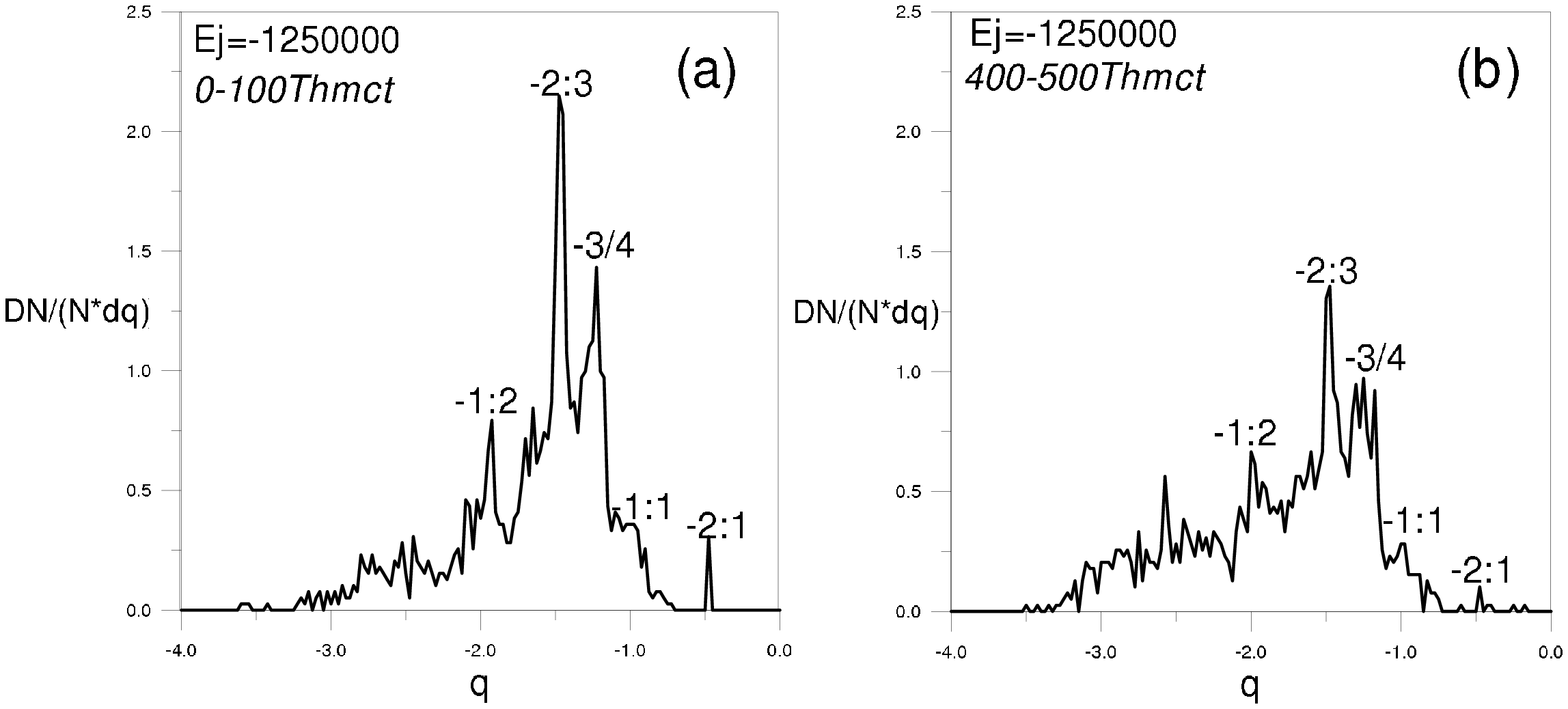}
\centering \caption{The distribution of the frequency ratios $q$ of
N-body chaotic orbits having initial conditions outside corotation
and belonging to the energy level $E_j=-1250000\pm 10000$, when the
orbits are integrated (a) for 100$T_{hmct}$ (corresponding to
$\approx$ 1/3 Hubble time) and (b) from 400$T_{hmct}$ to
500$T_{hmct}$ (corresponding to $\approx$ 1,5 Hubble time).}
\label{fig2}
\end{figure}
\begin{figure}
\centering
\includegraphics[width=8.cm,angle=-90] {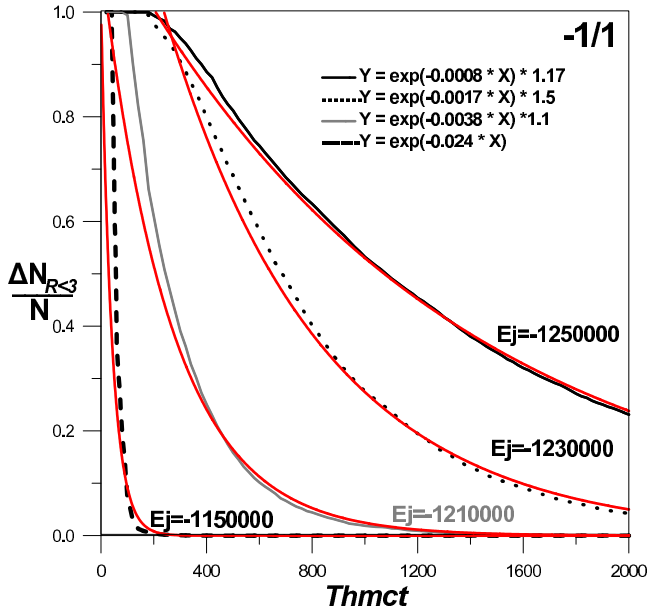}
\centering \caption{The percentage $\Delta N_{R<3}/N$ of the orbits
starting close to the $-1/1$ unstable periodic orbits that stay
inside $R=3r_{hm}$ as a function of time, in $T_{hmct}$, for four
different energy levels, together with the corresponding exponential
formulae \textbf{(red curves)}.} \label{fig2}
\end{figure}
In Figs. 7a,b the distribution of the $q$-values of these stars is
given, for the time intervals $0-100 T_{hmct}$ (Fig.$7a$) and
$400-500 T_{hmct}$ (Fig.$7b$). In this case there are no $PL_1,
PL_2$ orbits, but we notice some important resonances like $-2/1$
(outer Lindblad), $-1/1,-3/4,-2/3$ and $-1/2$. After one and a half
Hubble times (Fig.$7b$) the number of resonant stars has decreased
and many stars have escaped beyond the ends of the galaxy.
Nevertheless there is still an appreciable number of stars close to
these resonances. This is due to the stickiness phenomenon that
lasts for very long times, before the stars escape from the galaxy.
Comparing Figs. 6 and 7 we notice that in the case of chaotic orbits
that are restricted in the area outside corotation, stickiness to
resonances lasts for longer times in smaller values of Jacobi
constants than in greater ones. Thus in general, the diffusion of
chaotic orbits supporting the outer parts of the spiral arms
outwards is more slow than the diffusion of orbits supporting more
inner parts of the spiral structure. An example is given in Fig. 8
where the percentage of orbits starting close to the $-1/                                                               1$ unstable
periodic orbits that stay located inside a radius $R=3r_{hm}$ (which in fact confines the bound part of the galaxy, $r_{hm}$ being the
half mass radius of the system) is plotted as a function of time in
$T_{hmct}$, for four different energy levels (Jacobi constants),
namely for $E_j=-1250000$ (black solid curve), $E_j=-1230000$
(black dotted curve), $E_j=-1210000$ (gray curve) and  $E_j=-1150000$
(black dashed curve). After 1 Hubble time ($\approx 300 T_{hmct}$),
$95\%$ of the orbits with $E_j=-1250000$ is still located inside
$R=5r_{hm}$, while $\bf{90\%}$ of the orbits with $E_j=-1230000$, $40\%$ with
$E_j=-1210000$ and only $0.1\%$ of the orbits with $E_j=-1150000$ is
still located inside $R=3r_{hm}$. The functions $\Delta N_{R<3}$N
versus $T_{hmpt}$ for the various energy levels are approximated by
exponentials of the form
\begin{equation}
\Delta N _{R<3}/N=\alpha exp(-\lambda T_{hmpt})
\end{equation}
where $\alpha$ and $\lambda$ take the  values given in Table 1.\\
\begin{table}[ht]
\caption{The values of $\alpha$ and $\lambda$ in eq. (17)}
\centering
\begin{tabular}{c c c}
\hline\hline
E & $\alpha$ & $\lambda$ \\
\hline
-1250000 & 1.17 & 0.0008 \\
-1230000 & 1.50 & 0.0017 \\
-1210000 & 1.10 & 0.0038 \\
-1150000 & 1.00 & 0.0240 \\
\hline
\end{tabular}
\end{table}

The values of $\alpha$ are close to $\alpha=1$, while $\lambda$ can
be given by the approximate formula
\begin{equation}
\lambda=A exp(\Lambda E)
\end{equation}
with $A=1.8$ x $10^{15}$ and $\Lambda =3.4$ x $10^{-5}$. Therefore the diffusion is much
faster for larger Jacobi constants. On the other hand for smaller
Jacobi constants the diffusion is slower, and most of the stars
remain close to the outer parts of the spiral arms for more than a
Hubble time.

 In the case of chaotic orbits with initial
conditions inside corotation or close to it, the diffusion happens
quickly for an initial time interval corresponding to $\approx$ 1/3
of the Hubble time  (during which the spiral structure survives),
while later on it is very slow (see Fig. 24 of Harsoula et al.
2011). We therefore conclude that the outer parts of the spiral
structure of our N-body model survive for longer times than the
inner parts of the spiral structure.

\section{The role of asymptotic orbits}

\begin{figure}
\centering
\includegraphics[width=12.cm] {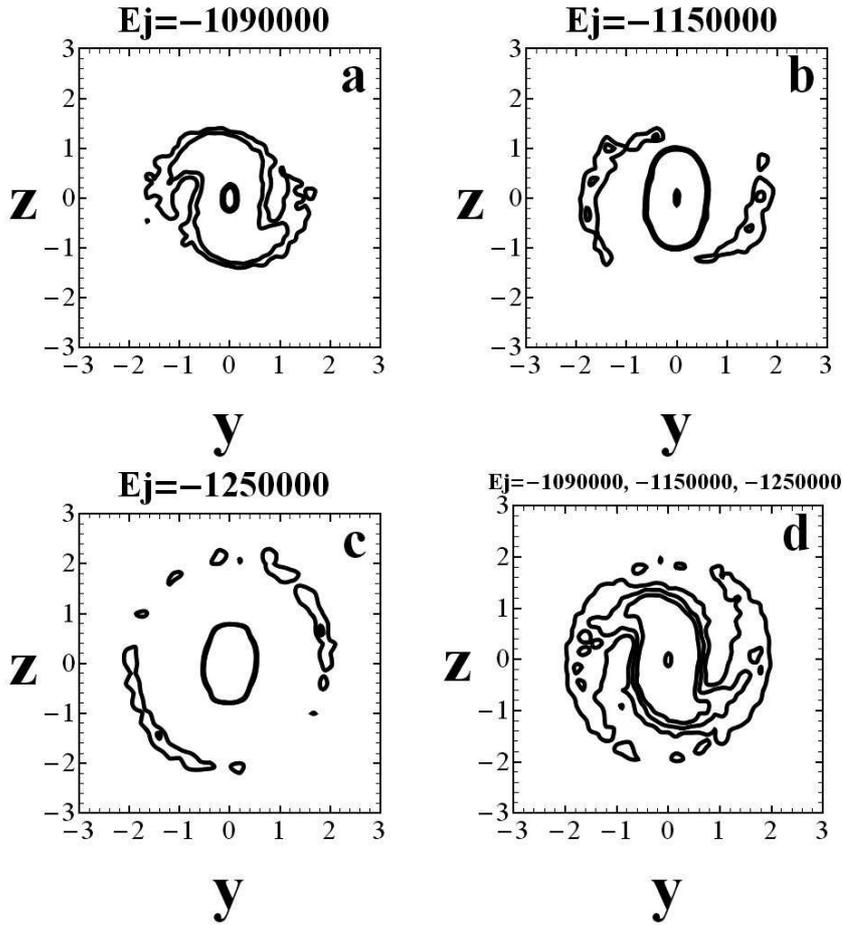}
\centering \caption{The isodensities of the N-body particles
belonging to three different energy levels, namely (a)
$E_j=-1090000\pm 10000$, where the areas inside and outside
corotation can communicate (b) $E_j=-1150000\pm 10000$, where the
areas inside and outside corotation cannot communicate (c)
$E_j=-1250000\pm 10000$ where again the areas inside and outside
corotation cannot communicate and (d) the isodensities of particles
belonging to all previous three levels. The spiral structure is
apparent here.} \label{fig2}
\end{figure}
In previous papers (Harsoula and Kalapotharakos 2009, Harsoula et
al. 2011) we have emphasized the role of stickiness of chaotic
orbits along the unstable asymptotic manifolds of the unstable
periodic orbits, in supporting the structure of the spiral arms. In
what follows, we investigate the role of the 2-D asymptotic orbits,
i.e. orbits having initial conditions on the unstable manifolds of
the various unstable periodic orbits, in supporting the spiral
structure of the model.
\begin{figure}
\centering
\includegraphics[width=12.cm] {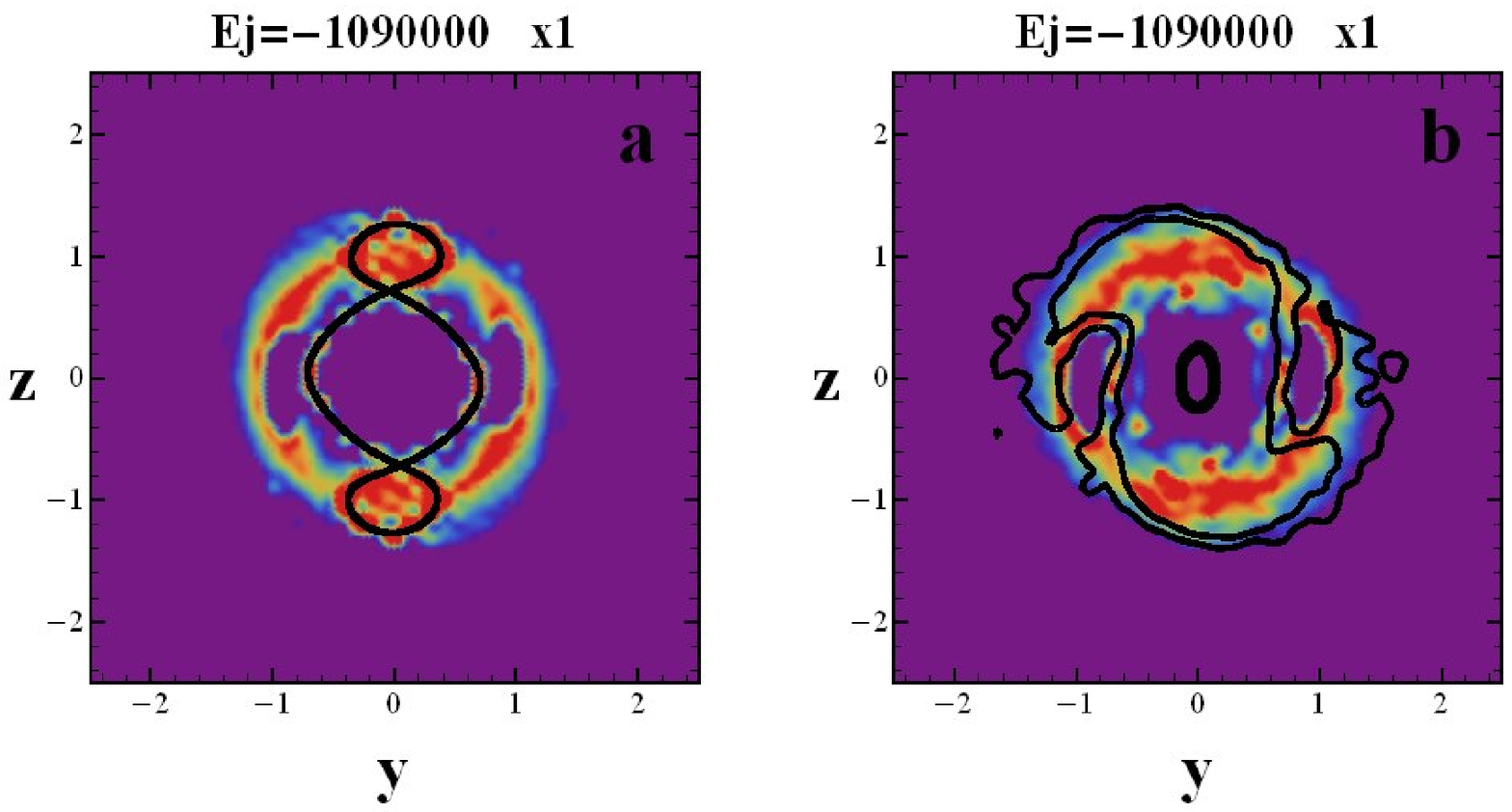}
\centering \caption{(a) The density distribution of 10000 2-D
asymptotic orbits (in color) belonging to the $2/1$ (or $x1$) family
having initial conditions inside corotation and Jacobi constant
$E_j=-1090000$. Superimposed is the orbit $x1$ plotted in black. (b)
The density distribution of 2-D orbits having initial conditions on
a grid close and around the unstable periodic orbit $2/1$ (in
color). Superimposed, in black, are the isodensities of the real
N-body particles belonging to the same energy level. The
corresponding time of integration of the orbits is $\approx$ one and
a half Hubble time.} \label{fig2}
\end{figure}
Orbits in different energy levels support different parts of the
spiral structure. This is obvious in Fig.9 where the isodensities of
the N-body particles are plotted on the configuration plane of
rotation, belonging to three different energy levels. More
precisely, particles having values of Jacobi constant $E_j=-1090000
\pm 10000$ correspond to the envelope of the bar and the innermost
parts of the spiral arms. Particles with values of Jacobi constant
$E_j=-1150000 \pm 10000$ correspond to parts of the spiral arms that
extend further beyond and finally particles with values of Jacobi
constant $E_j=-1250000 \pm 10000$ correspond to the outermost part
of the spiral arms. In Fig. 9d we plot the isodensities of particles
belonging to all previous energy levels of Figs. 9a,b,c. The spiral
structure of the galaxy is apparent.

In Harsoula et al. 2011 we studied the density distribution of $3-D$
orbits starting close to the various resonances. However we find
similar results if we consider the projections of the various orbits
on the plane of symmetry ($y-z$) of the galaxy (having the bar along
the z-axis).

In Fig. 10 an example of the density distribution of 2-D asymptotic
orbits of an unstable periodic orbit is plotted (in color) for a
Jacobi constant $E_j=-1090000$, where the areas inside and outside
corotation can communicate. More precisely in Fig.10a we plot (in
color) 10000 asymptotic orbits of the $2/1$ (or $x1$) family, having
initial condition inside corotation, together with the unstable
periodic orbit (in black). The corresponding time of integration of
the orbits is $\approx$ one and a half Hubble time. For an initial
time interval equal to $\approx$ 1/5 of the Hubble time, the chaotic
orbits stay close to the periodic orbit, following its shape, but
later on they are diffused outwards modulating the inner parts of
the spiral structure. If we take the same number of orbits, with
initial conditions not on the unstable manifold, but on a grid close
and around the unstable periodic orbit having small deviations from
it, on the ($z,\dot{z}$) surface of section(Fig. 10b), we find that
the distribution of the orbits (in color) follows the inner parts of
the spiral pattern, derived from the isodensities of the real N-body
particles of the corresponding energy level (in black). This is an
example of stickiness of chaotic orbits along the unstable
asymptotic manifolds of the unstable periodic orbits. In fact the
stickiness of chaotic orbits delays their diffusion outwards and is
responsible for the survival of the spiral structure of the galaxy
for more that 10 rotations of the bar (see Harsoula et al.
2011).

Similar results are found for the $PL_1,PL_2$ orbits near $L_1,L_2$
at the end of the bar, for the $-1/1$ orbits outside corotation and
for the $PL_4,PL_5$ orbits around $L_4$ and $L_5$.

\begin{figure}
\centering
\includegraphics[width=8.cm] {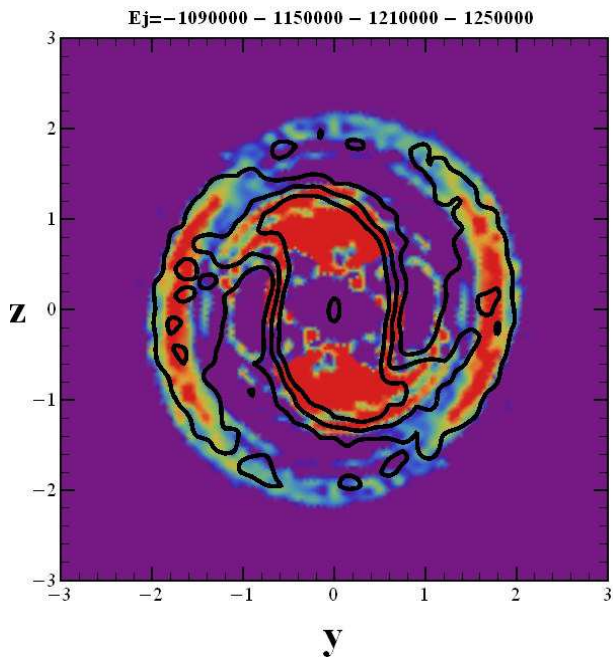}
\centering \caption{The density distribution of the orbits
starting close to the resonances $2/1, PL_1, PL_2, -1/1$ and $-2/1$ for
various Jacobi constants, superimposed with the isodensities of the
real N-body particles belonging to the corresponding energy levels
(black curves).} \label{fig2}
\end{figure}

Finally, in Fig. 11 we present the density distribution of  the
sticky chaotic orbits near the resonances $2/1,PL_1, PL_2,-1/1,-2/1$
superimposed with the isodensities of the real N-body particles
belonging to the corresponding energy levels. We conclude that by
using a sample of sticky chaotic orbits around a number of unstable
periodic orbits inside and outside corotation in different energy
levels, we are able to reproduce quite well the outer envelope of
the bar and the spiral structure of the galaxy.

\section{Conclusions}

The main conclusions of our paper are the following:

1) Stickiness of chaotic orbits close to the unstable asymptotic
manifolds of various periodic orbits  delays the diffusion of these
orbits outwards and therefore modulates the shape of the spiral
structure of the galaxy for more than 10 rotations of the bar,
corresponding to 1/3 of the Hubble time.

2) Chaotic orbits that are limited outside corotation modulate the
outer parts of the spiral structure for smaller values of Jacobi
constant while orbits with greater values of Jacobi constant
modulate the inner parts of the spiral structure. Moreover, in our
N-body model, stickiness to resonances for smaller values of Jacobi
constants lasts for longer times than stickiness for greater values
of Jacobi constants.

3) Asymptotic orbits (having initial conditions on the unstable
asymptotic curve of an unstable periodic orbit) stay located close
to the periodic orbit for an initial interval of time, following the
shape of this specific orbit, before diffusing from it and
supporting the spiral structure. Chaotic orbits having initial
conditions inside corotation modulate the envelope of the bar and
the innermost spiral structure during a time interval of fast
diffusion ($\approx 1/3$ of the Hubble time) and then they are
diffused outwards with  much slower rates.

4) Using a sample of sticky chaotic orbits close to  a number of
unstable periodic orbits inside and outside corotation, in different
energy levels, we are able to reproduce quite well the outer
envelope of the bar and the spiral structure of the galaxy.

\end{document}